\documentclass[preprint,aps,showpacs]{revtex4}

\usepackage{graphicx}

\usepackage{bm}% small greeck bold}

\begin{document}

\title{Relativistic particle
incoherent scattering in oriented crystals}

\author{Victor V. Tikhomirov}

\affiliation{Research Institute for Nuclear Problems, Belarusian
State University, Minsk, Belarus}

\date{\today}

\begin{abstract}

The coherent process of particle  deflection by aligned atomic
strings and planes of oriented crystals is accompanied by the
incoherent scattering by atomic cores. While the coherent particle
deflection, described by the axial or planar averaged potential,
becomes more and more classical, the incoherent scattering remains
essentially quantum at relativistic energies. Though the latter
reminds the scattering by atoms of amorphous medium at high enough
momentum transfers, at the smallest ones the incoherent scattering
process in crystals experiences some modification by the influence
of the inhomogeneity of the atom distribution in the plane, normal
to the crystal axis or plane. Considering the axial case as a more
general example, we present a consistent theory of high energy
particles incoherent scattering in oriented crystals. The latter
takes into consideration both the quantum scattering nature and
the atom distribution inhomogeneity, revealing a limited
applicability of the scattering cross section notion. A way to
incorporate the quantum scattering features into the widely used
classical trajectory simulations is elaborated using a specific
mean square scattering angle definition.
\end{abstract}


\maketitle

\section{Introduction}
High energy particle interaction with oriented crystals makes it
possible both to observe many remarkable phenomena and apply them
to develop diverse sources of x- and gamma-radiation
\cite{bar,akh,bai}, to efficiently deflect high energy particle
beams \cite{tsy,bir,sca}, to measure and even to modify elementary
particle properties, such as magnetic momenta \cite{bar}, to
reduce the thickness of particle detectors as well as to make the
latter sensitive to both direction and polarization
\cite{bas,bar2,ban,ban2}. All the pronounced effects, induced by
the coherent particle interaction with oriented crystal lattice,
are described by the averaged (continuum) potential of atomic
strings/planes \cite{lin,tho} introduced by J. Lindhard, who also
proved that the particle motion in the averaged potential can be
treated classically at high enough energy.

Despite the large strength of the coherent effects in crystals,
all their applications are essentially limited by the incoherent
scattering effects relational, but not completely similar to the
scattering process in amorphous media or randomly oriented
crystals. Most severely incoherent scattering by nuclei limits the
deflection of negatively charged particles by bent crystals
\cite{maz,syt} as well as both the channeling \cite{bar,akh,bai}
and crystal undulator \cite{bar,kor,bel,bar3} radiation of the
same. That is why, to consider any application of the coherent
effects in crystals, it is mandatory both to understand and
properly treat the incoherent ones. However neither a consistent
theory, nor a commonly recognized view on the nature of high
energy particle incoherent scattering still exist.

Channeling effect study began from the classical particle motion
simulations by binary collision method \cite{rob}. Being correct
for MeV-energy ions, the latter is inapplicable in relativistic
case, as, following \cite{lin, bor, lan, lin2}, we remind in
Section 2. This way both a fundamental problem and practical
necessity of treating quantum effects in the incoherent scattering
of high energy particles, moving along classical trajectories in
the averaged crystal potential, arise. Following \cite{kit}, the
multiple scattering theory in homogeneous medium was initially
applied \cite{tar} to sample the angles of classically moving
particle scattering on the atomic planes. However this approach
was not completely satisfactory since the plane/string atomic
density is strongly inhomogeneous at the scale of $u_1 < 0.1 $\AA,
which makes it inadequate to use a fixed density value for impact
parameters $b$ from the interval $u_1 \leq b \leq R$, where 0.1
\AA $ < R \leq 0.5 $\AA~ are atomic radii.

The recently observed \cite{maz2} influence of crystal atom
density inhomogeneity on incoherent particle scattering was
predicted along with the development of coherent bremsstrahlung
theory \cite{ter}, which also described the incoherent radiation
and pair production reduction, caused by the same of incoherent
scattering. Being developed in the quantum nature plane wave
\cite{ter} and reproduced in the classical straight-line
approximations \cite{akh, bai}, coherent bremsstrahlung theory
remains valid at particle incident angles, at least, a few times
exceeding the critical channeling angle, involving only the
incoherent scattering intensity averaged over the uniform particle
flux implied by both approximations.

However, at channeling and close to channeling conditions,
particles spend different time at different locations in the plane
of transverse motion or even do not reach some of them at all. The
uniform flux approximation, accordingly, loses its applicability,
making necessary to describe incoherent particle scattering at
each point individually, taking into consideration the scattering
by nuclei at the far distances $R > u_1$ from the trajectory. This
problem, in principle, is solved by the full quantum treatment of
transverse particle motion \cite{baz, baz2}, which, in fact, is
both really necessary and practically feasible only at the
electron and positron energies of a few dozen MeV and less.
However most of the current investigations are conducted at GeV
\cite{maz, syt, ban3} and higher \cite{bar, akh, bai, tsy, bir,
sca, bas, bar2, ban, ban2, kor, bel, bar3, tik, tik2, tik3,
art,tik4, bar4, gui} energies, at which the quantum description of
particle motion in the averaged crystal field becomes both
redundant and cumbersome, revealing the necessity of the
introduction of classical particle motion features into the
treatment of their incoherent scattering by the inhomogeneously
distributed nuclei. Till now, a redefinition of the scattering
impact parameter upper limit  $R \rightarrow u_1$ \cite{lju, tik,
tik2, tik3, art} has been used, which did not take into
consideration the incoherent scattering dependence on transverse
particle coordinate, giving, thus, mostly a qualitative estimate.

To develop a really quantitative approach, the Wigner function of
the transverse particle motion phase space \cite{ish} is applied
for the local treatment of the incoherent scattering in Section 3,
in which an essentially novel formula for local probability of
incoherent scattering of a classically moving high energy particle
is derived in the axial case. A method of consistent inclusion of
the quantum scattering features into the simulations of
relativistic particle classical motion in the averaged atomic
string/plane potential is detailed in Section 4. We reveal, that
the strong enough inhomogeneity of the string atom nuclei
distribution in the plane of transverse particle motion results in
impossibility to introduce a local scattering probability for the
small angles and, to preserve the classical trajectory
simulations, suggest to apply the newly introduced mean scattering
angles. Considerable attention is also payed both to the
interrelation and individual roles of single and multiple
scattering processes, quite differently treated for decades.

This Preprint presents the revised version of the paper
\cite{tik6}, both corrected and simplified according to Erratum
\cite{tik7} in order to make the developed approach to the
high-energy particle incoherent scattering in oriented crystals
more exact and available for implementation into simulations.

\begin{figure} %1
\label{Fig1}
 \begin{center}
\resizebox{80mm}{!}{\includegraphics{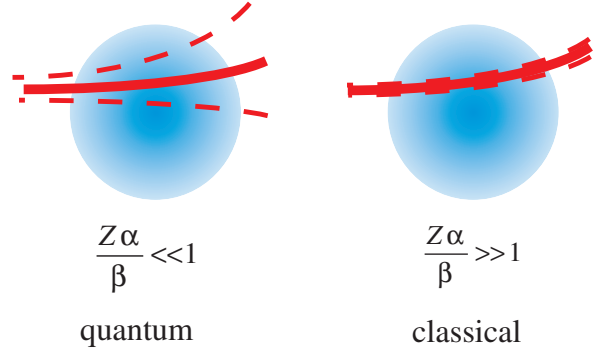}}\\
\caption{Comparison of diffractive wave packet divergence (dashed)
and classical particle deflection (solid line) in quantum (left)
and classical cases.}
\end{center}
\end{figure}

\section{Quantum nature of relativistic particle single atom scattering}

First of all, remind, why, on the contrary to the coherent, the
incoherent scattering becomes quantum at relativistic energies
\cite{lin,bor,lan,lin2}. The inapplicability of classical
mechanics to relativistic elementary (with a unit charge $|z| =
1$) particle scattering by nuclei, in fact, directly follows from
a comparison of the quantum (diffractional) angular uncertainty
$\Delta \theta \sim \hbar/p b$, where $p$ is particle momentum and
$b$ impact parameter, with the classical deflection angle
$\theta_{cl} = 2 Z \alpha/p v b$, where $Z$ is atomic number,
$\alpha$ fine structure constant and $v = \beta c$ particle
velocity. Indeed, since $\Delta \theta >> \theta_{cl}$ at both $Z
\alpha / \beta << 1$ and $\beta \simeq 1$, the scattering angle
uncertainty exceeds the classical deflection angle, making the
trajectory notion inapplicable for relativistic elementary
particles, as Fig. 1 qualitatively illustrates. In other words,
while the usage of classical trajectories in the averaged crystal
potential becomes more and more justified with the energy increase
(above several dozen MeV for electrons and positrons) \cite{lin},
classical binary collision method \cite{rob}, on the opposite,
becomes inapplicable.

\begin{figure} %2
\label{Fig2}
 \begin{center}
\resizebox{80mm}{!}{\includegraphics{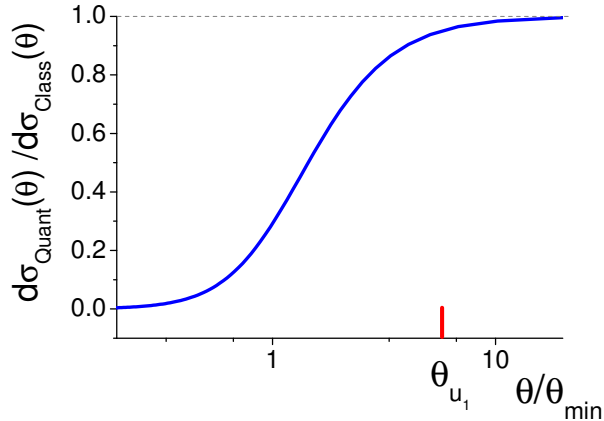}}\\
\caption{The ratio of "quantum" to "classical" cross sections vs
the scattering angle for Si atom ($\alpha Z =0.102$).}
\end{center}
\end{figure}

To illustrate the difference of quantum and classical predictions,
let us compare the corresponding cross sections, evaluated for
Yukawa atomic potential, characterized by the adopted from
\cite{mol,bet} and implemented into GEANT4 screening radius $R =
a_{TF}[1.13+3.76(\alpha Z/\beta )^2]^{-1/2}$, where $a_{TF}=0.8853
a_B Z^{-1/3}$ and $a_B$ are, respectively, Thomas-Fermi screening
and Bohr radii. Fig. 2 presents the angular dependence of the
ratio of "quantum" to "classical" cross sections, the latter of
which was evaluated following \cite{leh} in the relativistic case.
These cross sections, as is well known \cite{lan}, coincide at
large transverse momentum transfers $q >> \hbar/R$, corresponding
to the scattering angles $\theta >> \theta_{min} = \hbar/p R$.
However at $q \leq \hbar/R$ or $\theta \leq \theta_{min}$, when
the screening effect considerably modifies Coulomb potential,
classical approach overestimates \cite{jac} the scattering
intensity, demonstrating the loss of its applicability. Note, that
both the cross sections, compared in Fig. 2, have been calculated
under the assumption of a uniform incident particle flux, for
which classical mechanics overestimates the total cross section by
the factor $(2 Z \alpha)^{-1}$, reaching 25 for Si atom. At the
same time, on the opposite to the classical particle deflection in
binary collisions model, quantum mechanics can not be directly
applied to quantify the deflection of a particle moving along a
classical trajectory.

It will be shown below, that the correlations of particle
collisions with string atoms result in the incoherent scattering
reduction similar, in a sense, to a screening at momentum
transfers $q \leq \hbar/u_1$, corresponding to the scattering
angles $\theta_{u_1} \leq \hbar/p u_1$, where $u_1$ is the root
mean square amplitude of atom thermal vibrations. Since
$\theta_{u_1}$ usually two-three times exceeds $\theta_{min}$, one
can expect \cite{baz, lju} (see also \cite{tik, tik2, tik3, art})
that collision correlations have to result in an incoherent
scattering reduction which will be numerically described below,
taking for the first time into consideration its dependence on
transverse coordinates.

To have a natural measure of the difference of incoherent
scattering in crystals from the scattering in amorphous medium, we
will introduce a Kitagawa-Ohtsuki ansatz ("KO ansatz" below)
\begin {equation} %1
\displaystyle{\frac{d \Sigma_{KO}(\bm\rho)}{d\bm q}= \frac{4 Z^2
\alpha^2 n_n(\rho)}{v^2 [q^2+\kappa^2_{sc}]^2}},
\end{equation}
inspired by the paper \cite{kit} of M. Kitagawa and Y. H. Ohtsuki
to be equal to the product of the unperturbed microscopic
scattering cross section (relativistic cross section for the
screened Coulomb (Yukawa) potential here) by either planar or,
considered below as an example, axial
\begin {equation} %2
n_n(\rho) = \frac{\exp(-\rho^2/2 u^2_1)}{2  \pi u^2_1 d}
\end{equation}
nuclear number density, were $\rho$ is the distance from the
atomic string symmetry axis (see Fig. 3) and $d$ interatomic
distance in the string. One can mention that Eq. (1) reminds
macroscopic cross sections, widely used in the reactor physics, in
which the nuclear number density can also vary widely.

\begin{figure} %3
\label{Fig3}
 \begin{center}
\resizebox{50mm}{!}{\includegraphics{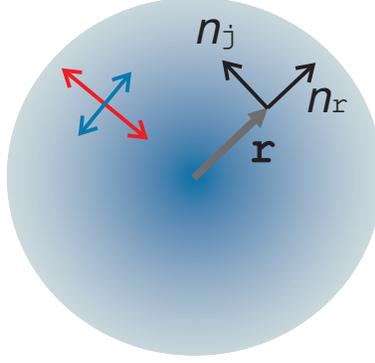}} \\
\caption{Transverse radius-vector $\bm \rho$ and the directions
$\bm n_{\bm\rho}$, $\bm n_{\bm\varphi}$ of extremal scattering
intensities at the background of the string field. The
double-sided arrows portray the excess of radial scattering over
the azimuthal one.}
\end{center}
\end{figure}

However, KO ansatz (1) combines a classical particle coordinate
$\rho$ with the quantum cross section only formally, neglecting
the nuclear number density (2) variation, which exceeds 100\% for
the largest substantial impact parameters $b \sim R > u_1$, as
well as missing both the effect of incoherent scattering reduction
by correlations \cite{maz2,ter,baz, lju, tik, tik2, tik3, art} and
azimuthal asymmetry of the same \cite{lju, art3}. To take the
influence of the nuclear density inhomogeneity into consideration,
we also introduce below some "macroscopic cross section", which,
formally reminding the same in reactor physics, possesses, in a
sense, a deeper meaning, combining in its general form the
inseparable characteristics of scattering probability and
scatterers' density distribution. At the same time, in the high
transfer momentum limit, the same macroscopic cross section
reduces to the product of nuclear density by a modified
microscopic cross section, preserving the effect of the density
(2) inhomogeneity and allowing one to incorporate quantum
scattering effects in classical trajectory simulations.

\section{Wigner function application to single atom scattering}

To introduce quantum features of incoherent scattering
consistently into the classical picture of high energy particle
motion in the averaged crystal potential, the Wigner function
aproach is most adequate \cite{ish}. Taking the axial case as an
example, let us consider the Wigner function
\begin {equation} %3
W(\bm\rho, \bm q)=\frac{1}{\pi^2}\int\psi^* (\bm\rho + \bm\chi)
\psi(\bm\rho - \bm\chi) \exp (2i \bm q \bm\chi)d^2 \chi,
\end{equation}
determined in the two-dimensional phase space $(\bm \rho,$ $\bm
q)$. $\bm \rho = (x, y)$, $\bm q = (q_x, q_y)$, of the impact
parameter plane nearly parallel to that of transverse motion of a
particle, scattered by the residual atomic potential \cite{baz,
baz2}
\begin {equation} %4
\delta U_{at}(\bm\rho, \bm\rho_n, z) =  U_{at}(\bm\rho, \bm\rho_n,
z-z_n) - \int U_{at}(\bm\rho, \bm\rho_n, z-z_n) n_n(\rho_n) d^2
\rho_n dz_n,
\end{equation}
emerging after the substraction of the averaged one, already taken
into consideration by the axial potential, which determines the
classical particle motion. The potential (4) is small and
localized enough to leave the three terms
\begin {equation} %5
\psi(\bm\rho, \bm\rho_n) = 1 - i \int_{at}\delta U_{at}(\bm\rho,
\bm\rho_n , z) \frac{dz}{v} - \frac{1}{2}\left[\int_{at}\delta
U_{at}(\bm\rho, \bm\rho_n,z) \frac{dz}{v}\right]^2
\end{equation}
of the eikonal wave function expansion, in which
\begin {equation} %6
\int_{at}\delta U_{at}(\bm\rho, \bm\rho_n, z) \frac{dz}{v}=
\frac{Z \alpha}{\pi v} \int \exp (i \bm\kappa \bm\rho) \frac{\exp
(- i \bm\kappa \bm\rho_n) - \exp ( - \kappa^2 u_1^2/2)}{\kappa^2 +
\kappa^2_s} d^2 \kappa.
\end{equation}
Assuming a purely classical particle motion in the averaged
potential both before and after the incoherent scattering at the
point $\bm\rho$ in transverse plane, we will evaluate the
distribution in momentum $\bm q$, transferred to the particle at
this point, being that of catenation of the classical trajectories
before and after the incoherent scattering. Substituting the wave
function product
\begin {equation} %7
\begin{array}{c}
\psi^* (\bm\rho + \bm\chi, \bm\rho_n) \psi(\bm\rho - \bm\chi,
\bm\rho_n) = 1 + i \int_{at}\delta U^*_{at}(\bm\rho + \bm\chi,
\bm\rho_n, z) \displaystyle{\frac{dz}{v}}  \\ - i \int_{at}\delta
U_{at}(\bm\rho - \bm\chi, \bm\rho_n, z)
\displaystyle{\frac{dz}{v}}+  \int_{at}\delta U^*_{at}(\bm\rho +
\bm\chi, \bm\rho_n, z) \displaystyle{\frac{dz}{v}} \int_{at}\delta
U_{at}(\bm\rho - \bm\chi, \bm\rho_n, z)
\displaystyle{\frac{dz}{v}}\\- \frac{1}{2} \int_{at}\delta
U^*_{at}(\bm\rho + \bm\chi, \bm\rho_n, z)
\displaystyle{\frac{dz}{v}}
\int_{at}\delta U_{at}(\bm\rho + \bm\chi, \bm\rho_n, z) \displaystyle{\frac{dz}{v}}\\
-  \frac{1}{2} \int_{at}\delta U^*_{at}(\bm\rho - \bm\chi,
\bm\rho_n, z) \displaystyle{\frac{dz}{v}} \int_{at}\delta
U_{at}(\bm\rho - \bm\chi, \bm\rho_n, z)
\displaystyle{\frac{dz}{v}}
\end{array}
\end{equation}
into Eq. (3) and assuming, that the unity in Eq. (7) corresponds
to the particle propagation in the absence of incoherent
scattering, one arrives to the "scattering" Wigner function
\begin {equation} %8
\begin{array}{c}
W_{\bm\rho_n}(\bm\rho, \bm q)= -\displaystyle{\frac{8 Z
\alpha}{\pi v}} \displaystyle{\left [\frac{\cos 2 \bm q (\bm\rho -
\bm\rho_n) - \cos (2 \bm q \bm\rho) \exp ( - 2 q^2 u_1^2)}{4 q^2 +
\kappa^2_s} \right]} \\ + \displaystyle{ \frac{4 Z^2 \alpha^2
}{\pi^2 v^2} \int \exp(2 i \bm\kappa \bm\rho) \left\{ \frac{\exp
[- i (\bm q + \bm\kappa ) \bm\rho_n] -  \exp [- (\bm q + \bm\kappa
)^2 u^2_1/2 ]}{(\bm q +
\bm\kappa)^2+\kappa^2_{sc}} \right\} }\\
\displaystyle{ \times\left\{\frac{\exp [i (\bm q - \bm\kappa )
\bm\rho_n] -  \exp [- (\bm q - \bm\kappa )^2 u^2_1/2 ]}{(\bm q -
\bm\kappa)^2+\kappa^2_{sc}} \right\} } d^2 \kappa \\ -
\displaystyle{ \frac{4 Z^2 \alpha^2 }{\pi^2 v^2} \exp(2 i \bm q
\bm\rho) \int \left\{ \frac{\exp [- i (\bm q + \bm\kappa )
\bm\rho_n] - \exp [- (\bm q + \bm\kappa )^2 u^2_1/2 ]}{(\bm q +
\bm\kappa)^2+\kappa^2_{sc}} \right\} }\\
\displaystyle{ \times\left\{\frac{\exp [-i(\bm q - \bm\kappa )
\bm\rho_n] -  \exp [- (\bm q - \bm\kappa )^2 u^2_1/2 ]}{(\bm q -
\bm\kappa)^2+\kappa^2_{sc}} \right\} } d^2 \kappa,
\end{array}
\end{equation}
describing particle incoherent scattering at the point $\bm\rho$
by an atom having a transverse radius vector $\bm\rho_n$. In
general, Eq. (8) can be applied to an arbitrary instant nuclear
distribution in thermal vibration coordinates $\bm\rho_n$, as is
discussed in Refs. \cite{art2, sol}. However, a probabilistic
interpretation of Eq. (8) does not look straightforward, since the
leading, linear in $\alpha$, term strongly oscillates between
positive and negative values. At the same time, the rest quadratic
terms in Eq. (8) reduce to the expected high momentum transfer
limit of the relativistic Rutherford (Mott) cross section
multiplied by delta function $\delta(\bm\rho - \bm\rho_n)$ at $q
u_1
>> 1$.

To simplify the problem realistically, we choose here the
traditional way \cite{ter, baz} of Eq. (8) convolution with the
scattering nuclei distribution (2), which nullifies the linear in
$\alpha$ contribution, leaving only the quadratic ones in (8).
Owing to the Fourier integral presence in Eq. (6), the double
integration in Eq. (3) results in the two-dimensional delta
function, trivializing the double integration over $\bm \kappa $
in one of the potentials (6), leaving the same in the another one
the sole integral in the resulting expression
\begin {equation} % 9
\begin{array}{c}
\displaystyle \frac{d \Sigma(\bm\rho)}{d\bm q}= \left\langle
W_{\bf{\rho}_n}(\bm\rho, \bm q) \right\rangle = \frac{4 Z^2
\alpha^2 }{\pi^2 v^2}\frac{1}{d} \Biggl\{ \int \cos(2\bm\kappa
\bm\rho) \frac{\exp(-2\kappa^2 u_1^2) - \exp[-(q^2 +
\kappa^2)u_1^2]}{[(\bm q +\bm\kappa)^2+\kappa^2_{sc}] [(\bm q -
\bm\kappa)^2+\kappa^2_{sc}]}d^2 \kappa \\
 -\cos(2\bm q \bm\rho) {\displaystyle \left[ \frac{ \pi \ln
(\sqrt{q^2/\kappa^2_{sc}+1} + q/ \kappa_{sc})}{q
\sqrt{q^2+\kappa^2_{sc}}} \exp[-2 q^2 u_1^2] \right. } \\
{\displaystyle \left. - \int \frac{\exp[-(q^2 + \kappa^2)u_1^2]
~d^2 \kappa}{[(\bm q +\bm\kappa)^2+\kappa^2_{sc}] [(\bm q -
\bm\kappa)^2+\kappa^2_{sc}] } \right] }\Biggr\}.
\end{array}
\end{equation}
for Wigner function, for which, just renaming the integration
variables, the unitarity condition $\int d^2 q \left\langle
W_{\bf{\rho}_n} (\bf{\rho, q})\right\rangle =0$ can be readily
checked. Eq. (9) can be widely used for the quantum treatment of
incoherent scattering of high energy particles, classically moving
in the average crystal potential.

As will be shown in Section 4, Eq. (9) predicts quite peculiar
behavior at momentum transfers $q < \hbar/u_1$, quickly
approaching at $q > 2~ \hbar/u_1$ its asymptote
\begin {equation} % 10
\begin{array}{c}
\displaystyle{\frac{d \sigma_{mod}(\bm\rho)}{d\bm q} =
\frac{1}{n_n(\bm\rho)}\frac{d \Sigma(\bm\rho)}{d\bm q} }\\
\displaystyle{ \simeq \frac{d \sigma_{Ruth}(\bm q_)}{d\bm q}
\left\{ 1 - \frac{\rho^2}{2 q^2 u_1^4} \left(1-8e^{-\rho^2/2 u_1^2
- q^2 u_1^2}\right)\left[\frac{2(\bm q \bm \rho)^2}{q^2 \rho^2} -
1\right] \right. } \\ \displaystyle{ \left. + 2 \pi^2 \cos(2\bm q
\bm\rho) \left[e^{-q^2 u_1^2} - q^2 u_1^2 e^{-2q^2 u_1^2} \ln
\frac{2 q}{\kappa_{sc}}\right] e^{\rho^2/2 u_1^2} \right\} },
\end{array}
\end{equation}
which will be used to clarify the nature of the considered
effects. At still larger momenta $q
>> \hbar/u_1$, corresponding to the impact parameters $b \sim
\hbar/q << u_1$, at which the influence of the nuclear
distribution inhomogeneity on scattering process must vanish, Eq.
(9) reduces to the product
\begin {equation} % 11
\displaystyle{ \left(\frac{d \Sigma(\bm\rho)}{d \bm q} \right)_{q
>> \hbar/u_1}}
 \rightarrow \frac{4 Z^2 \alpha^2 n_n(\rho)}{q^4} = \displaystyle{\frac{d \sigma_{Ruth}}{d\bm
q}} n_n(\rho)
\end{equation}
of Rutherford cross section by the nuclear density. Note, that
this "natural" limit takes place owing to both the residual
incoherent scattering potential (4) introduction and averaging
over the nuclei distribution (2).

\section{Quantum incoherent scattering features and their incorporation into classical
trajectory simulations}

\subsection{Local scattering probability and mean square angles}

Since both Eq. (9) and (10) demonstrate a strong scattering
asymmetry, also emphasized in \cite{lju,art3}, we will consider
Wigner function (9) properties for both radial $\bm q = (q_\rho,
0)$ and azimuthal $\bm q = (0, q_\varphi)$ transferred momenta
(see Fig. 3). Fig. 4 illustrates the diversity of the low-momentum
(9) behavior at different radial coordinates, corresponding to
both various nuclear densities and relative numbers - see Fig. 5.
The incoherent scattering modification considerably affects, in
fact, solely the smallest momentum transfers $q \lesssim
\hbar/u_1$, only $R/u_1 \div 2-3$ times exceeding the minimal
effective angle $1/Rp$ of particle scattering by a single atom. At
such small momentum transfers the incoherent scattering
differential probability drops below the normalization value,
demonstrating the effect of incoherent scattering reduction by
collision correlations \cite{maz2,ter, baz, lju, art, tik}, and
even becomes negative.

\begin{figure} %4
\label{Fig4}
 \begin{center}
\resizebox{60mm}{!}{\includegraphics{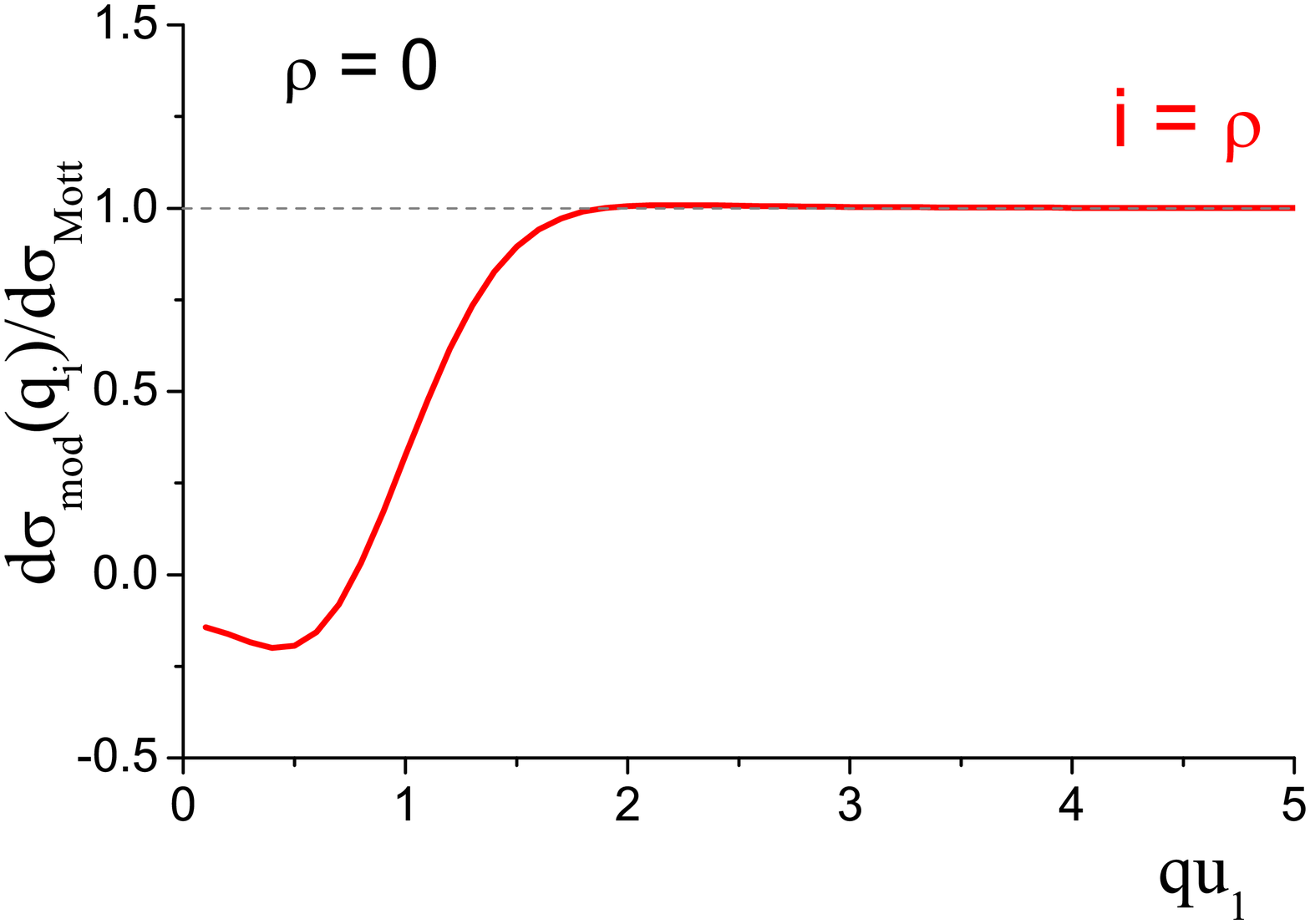}}
\hspace{1cm}\resizebox{60mm}{!}{\includegraphics{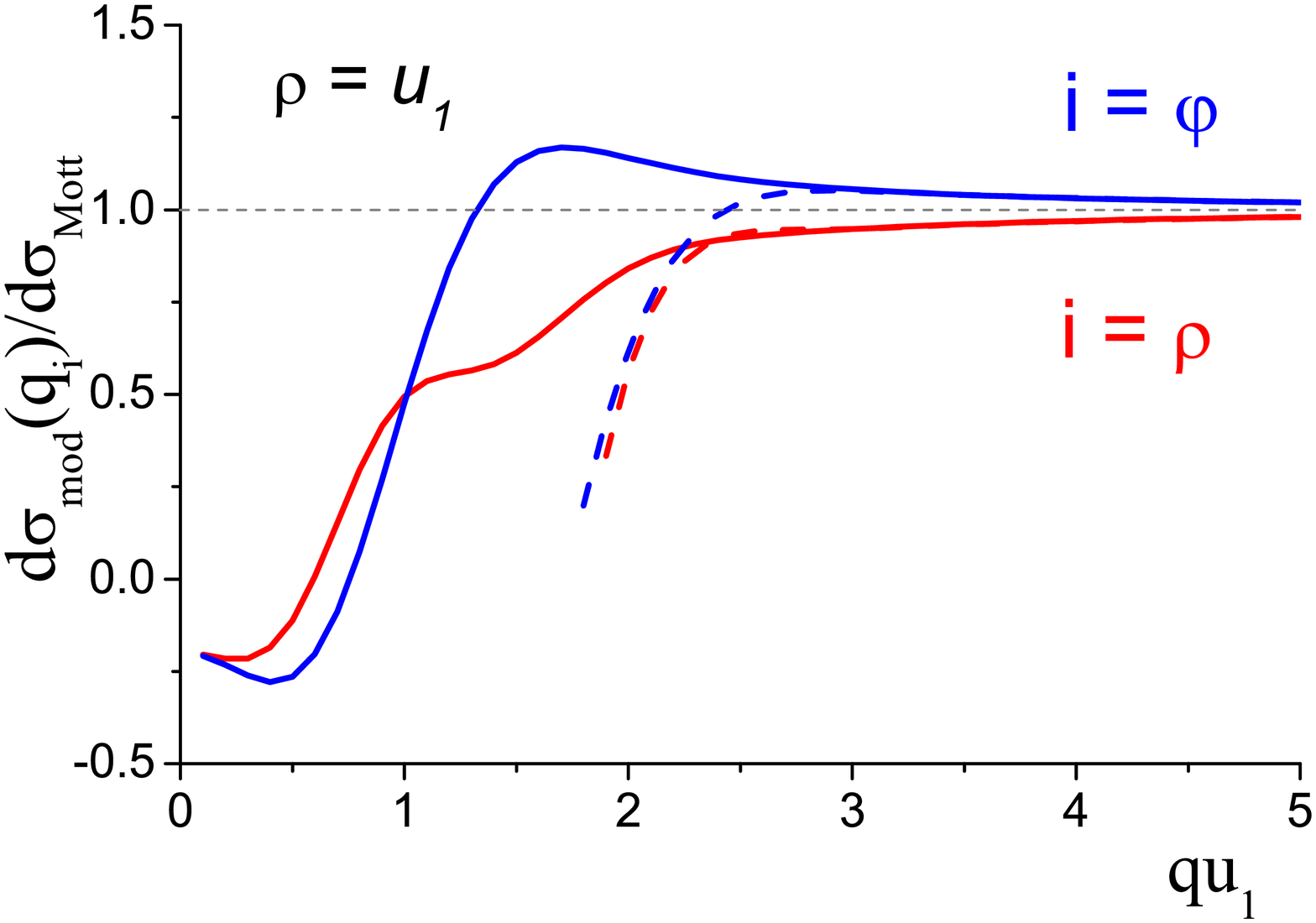}} \\
\resizebox{60mm}{!}{\includegraphics{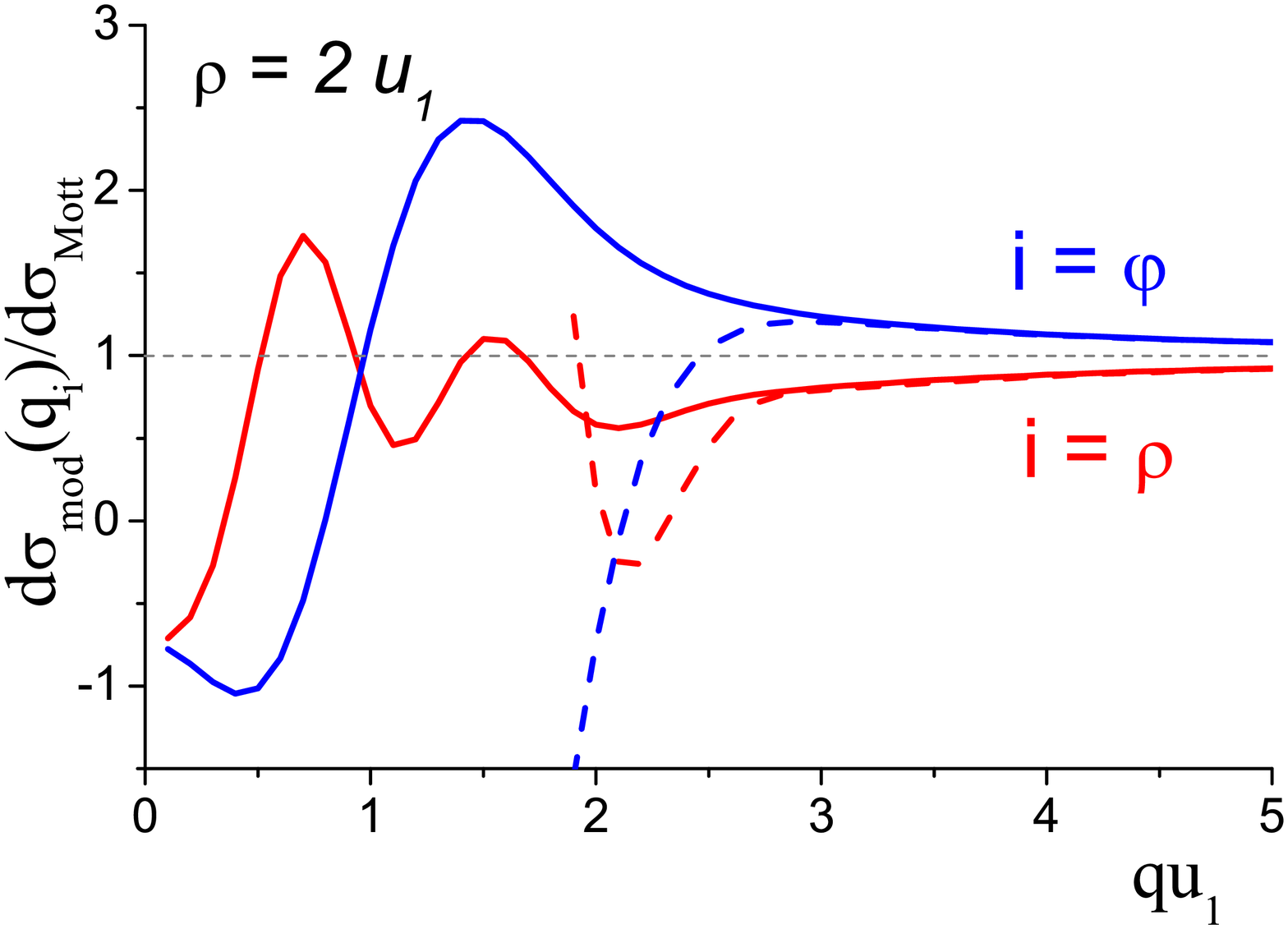}} \hspace{1cm}
\resizebox{60mm}{!}{\includegraphics{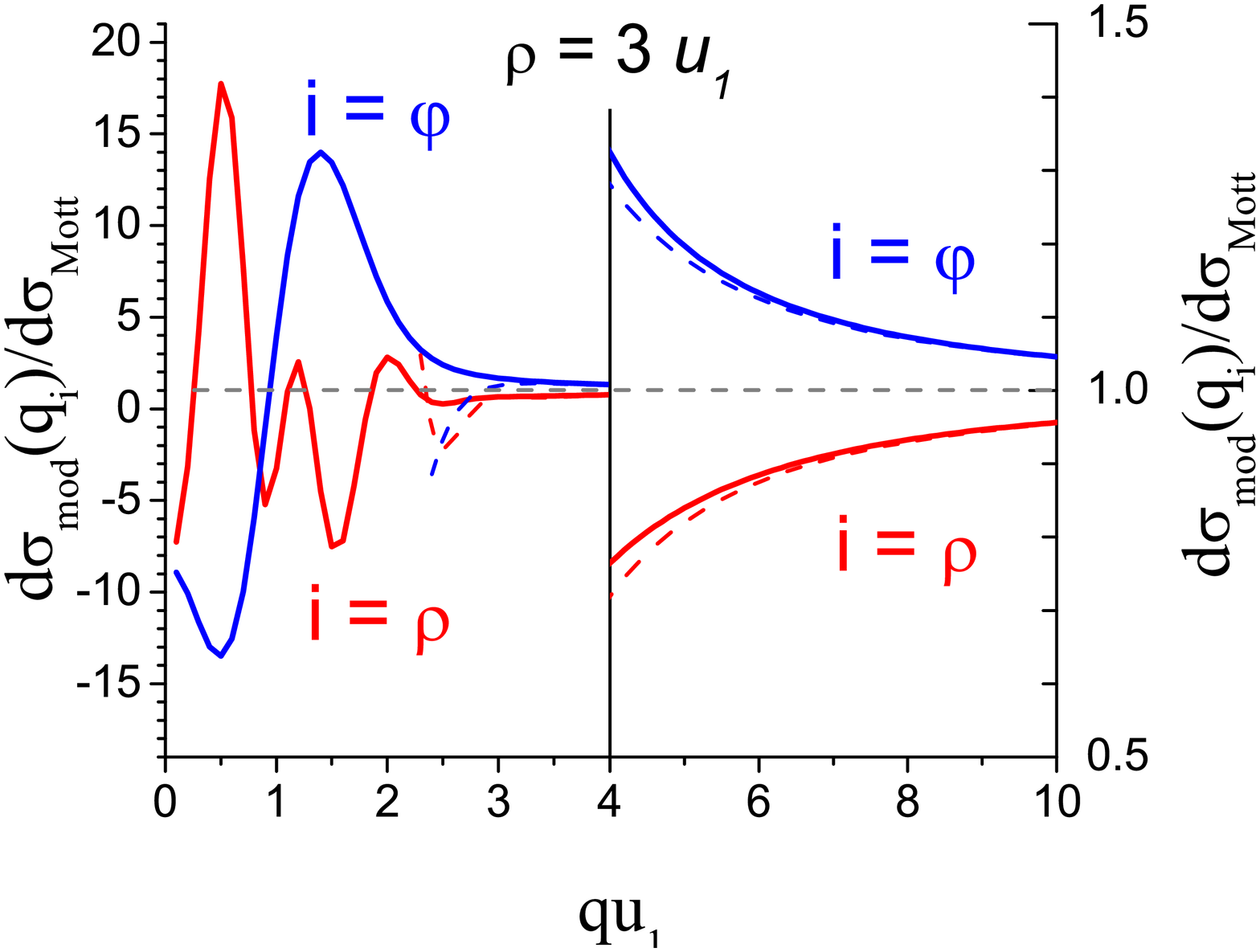}}\\
\caption{Modified cross section dependence on momentum transfer
for radial ($i = \rho$) and azimuthal ($i = \varphi$) scattering
directions, evaluated using the general (9) (solid) and asymptotic
(10) (dashed lines) formulae for the radial coordinates $\rho = 0$
(top left), $\rho = u_1$ (top right), $\rho = 2 u_1$ (bottom left)
and $\rho = 3 u_1$ (bottom right) of the scattering point. }
\end{center}
\end{figure}

\begin{figure} %5
\label{Fig5}
 \begin{center}
\resizebox{70mm}{!}{\includegraphics{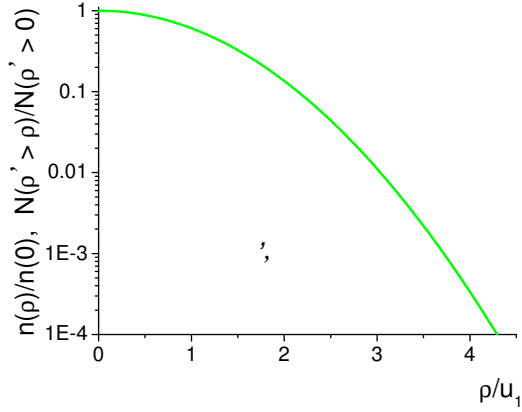}} \\
\caption{Radial dependence of both the nuclear number density (2)
and, proportional to the latter, number of the nuclei $N(\rho'
>\rho)$, situated at the radial distances $\rho' > \rho$, exceeding
the plotted $\rho$ value, both measured in units of their maximum
values reached at $\rho = 0$.}
\end{center}
\end{figure}

A capability to attain negative values is an essential property of
Wigner function, reflecting its quantum nature. The Wigner
function negativity does not allow one to unconditionally treat it
as a scattering probability. To retain, nevertheless, the
possibility to simulate quantum incoherent scattering of
classically moving particles, we put forward here a more indirect
way, resembling formally the "multiple scattering approach",
applied here, however, in the absence of positively determined
scattering probability (cross section). The point is that, since
Rutherford cross section peaks at the lowest q, the small angle
particle deflection often manifests itself as a multiple
scattering process, characterized by the mean square scattering
angle per unit length \cite{mol,bet,jac}. One can, similarly, use
Eq. (9) to introduce the same for both radial $i=\rho$ and
azimuthal $i=\varphi$ scattering directions
\begin {equation} % 12
\displaystyle{\frac{d \theta^2_i (\bm\rho)}{d z} =
\int_{q<q_{max}^{MS}} \frac{d \Sigma(\bm\rho)}{d\bm q}
\frac{q^2_i}{p^2} d^2q}.
\end{equation}
Fig. 6 illustrates the dependence of the latter on the momentum
integration limit $q_{max}^{MS}$, demonstrating again the drastic
scattering asymmetry, which can be further taken into
consideration by making $q_{max}^{MS}$ dependent on both radial
coordinate and transferred momentum direction. Since the cross
section approaches the Rutherford $q^{-4}$-type behavior at large
q, for which the values (12) are definitely positive, one can
expect that large enough $q_{max}^{MS}$ value will assure (12)
positivity, making possible routine scattering angle sampling.
Suggesting here to use the positively determined mean square
angles (12), we have in mind that, despite the absence of the
local probabilistic interpretation, in general, Wigner function is
applicable to evaluate the consistent integral values of both
scattering probability and mean square scattering angle,
validating suggested sampling method. Note also that the
complicated by the distant interference effects quite large Eq.
(9) contribution to the Eq. (12) integrand is, in fact, strongly
suppressed by the $q^3$ factor.
\begin{figure} %6
\label{Fig6}
 \begin{center}
\resizebox{60mm}{!}{\includegraphics{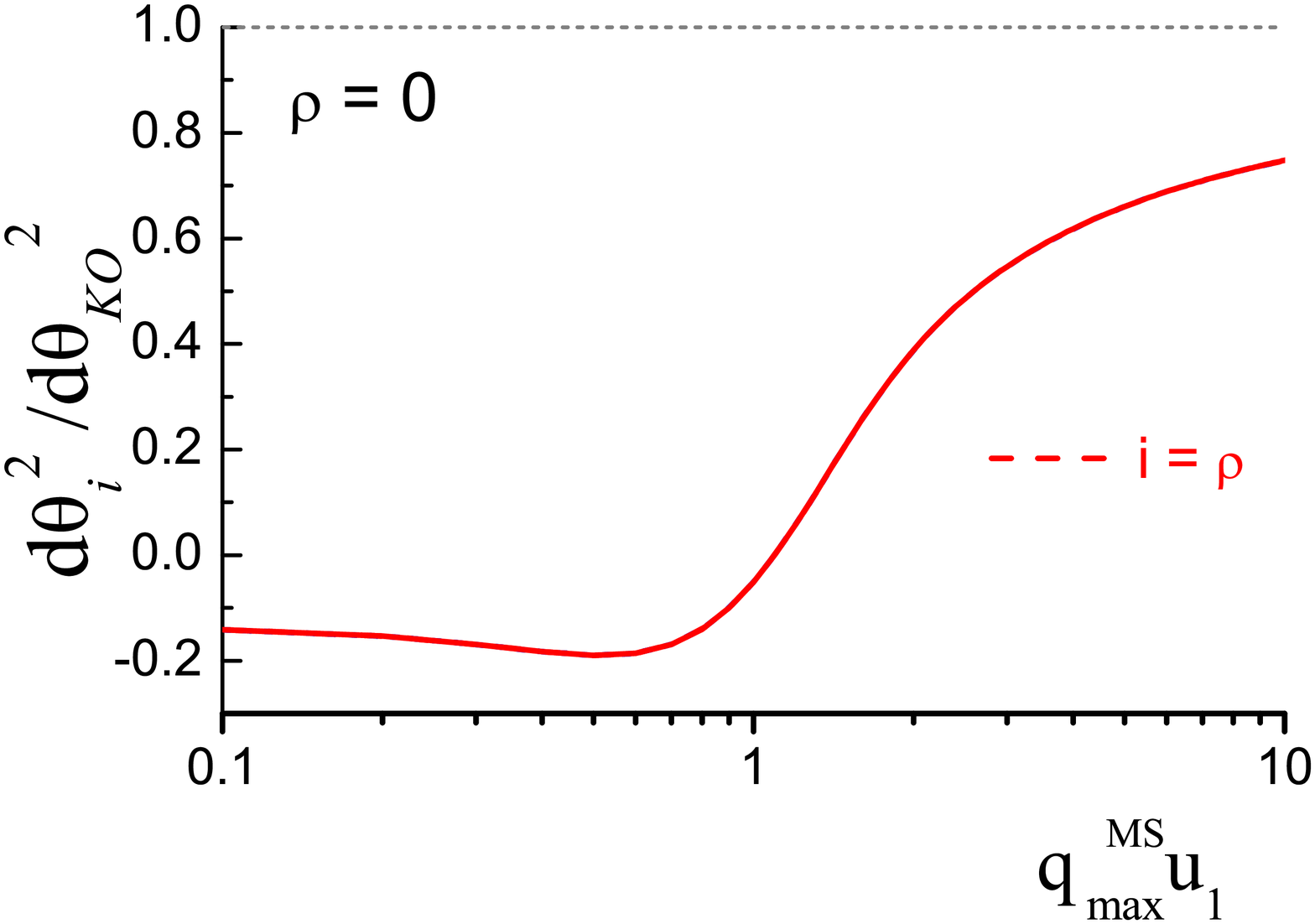}}
\hspace{1cm}\resizebox{60mm}{!}{\includegraphics{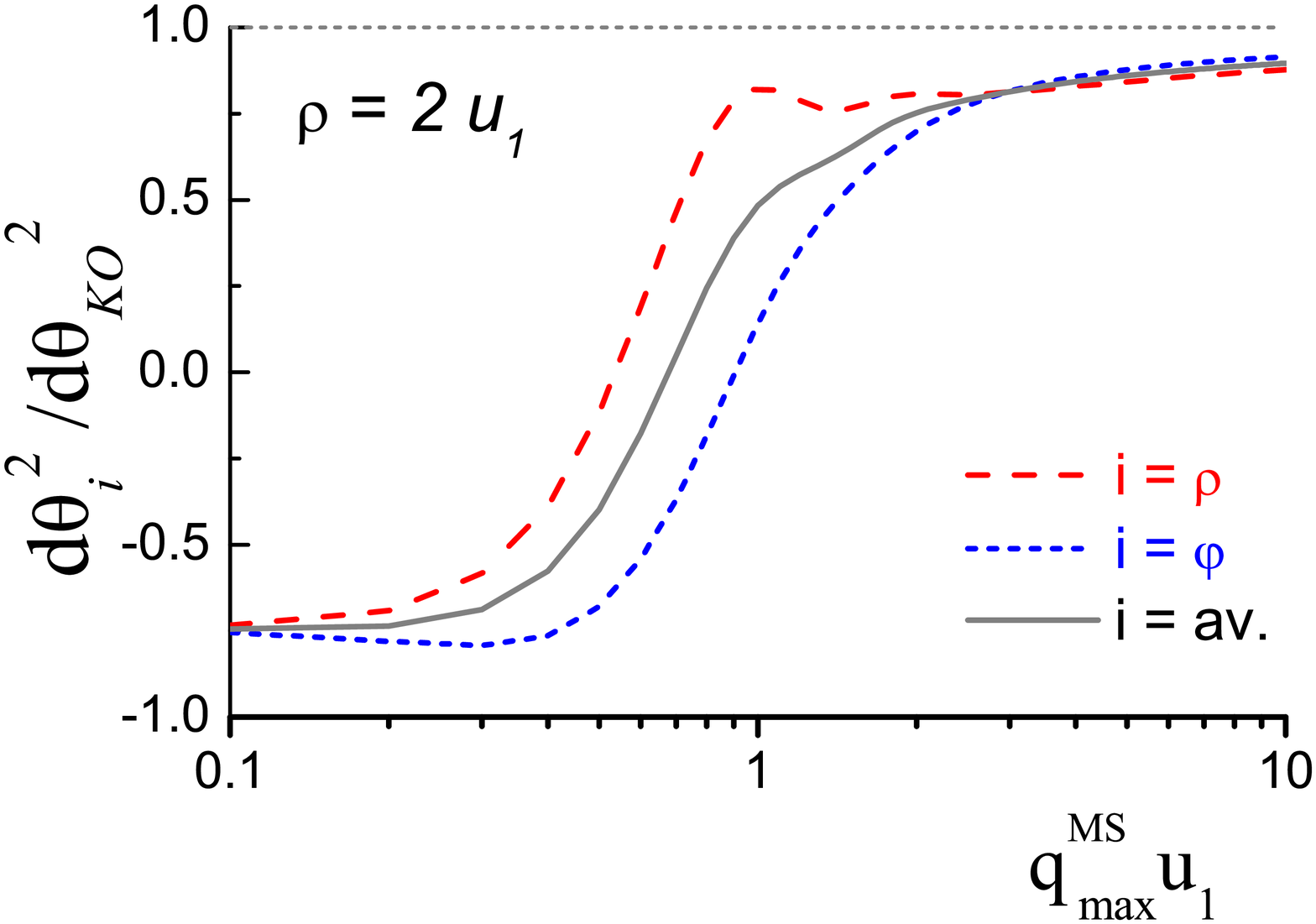}} \\
\resizebox{60mm}{!}{\includegraphics{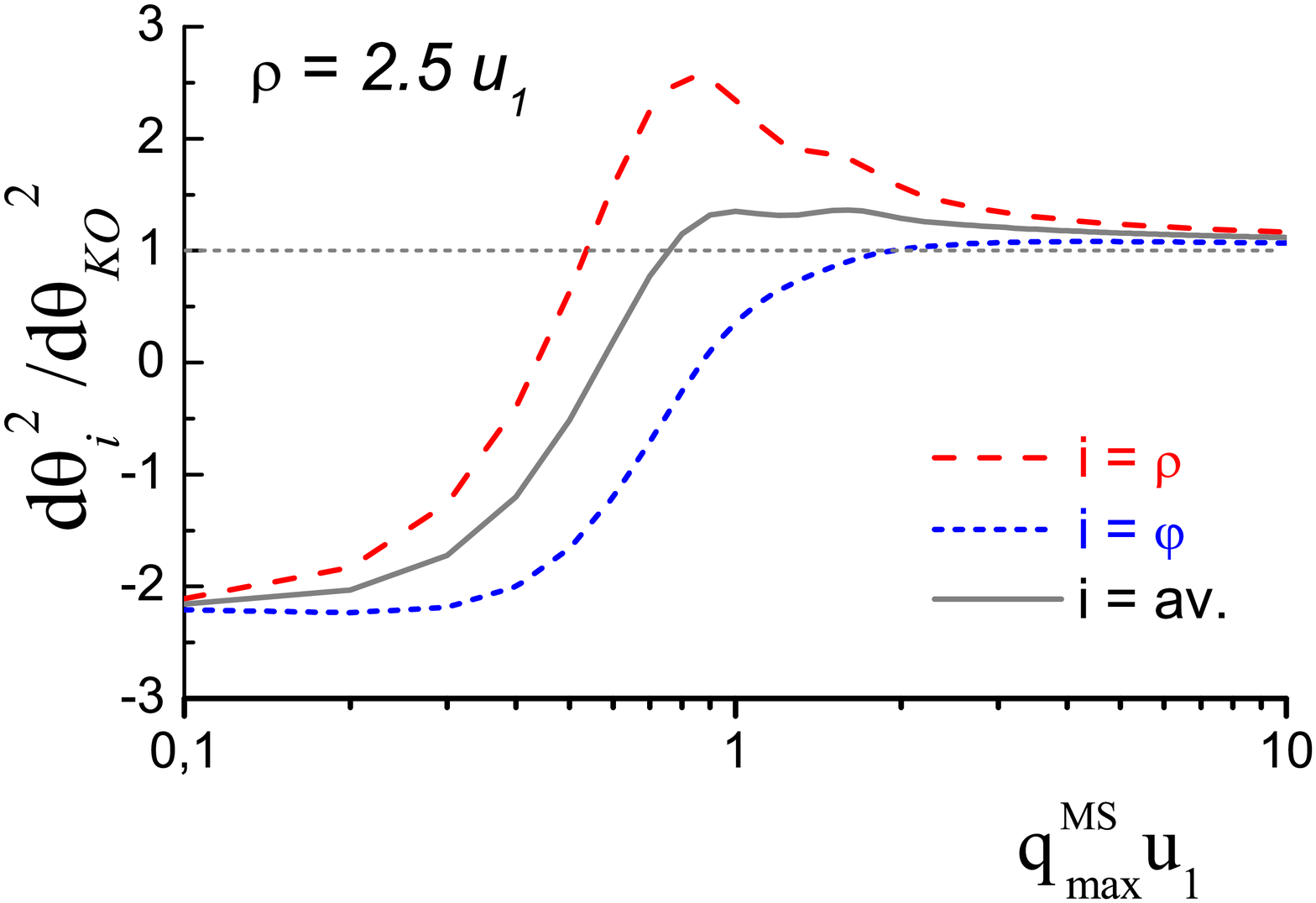}} \hspace{1cm}
\resizebox{60mm}{!}{\includegraphics{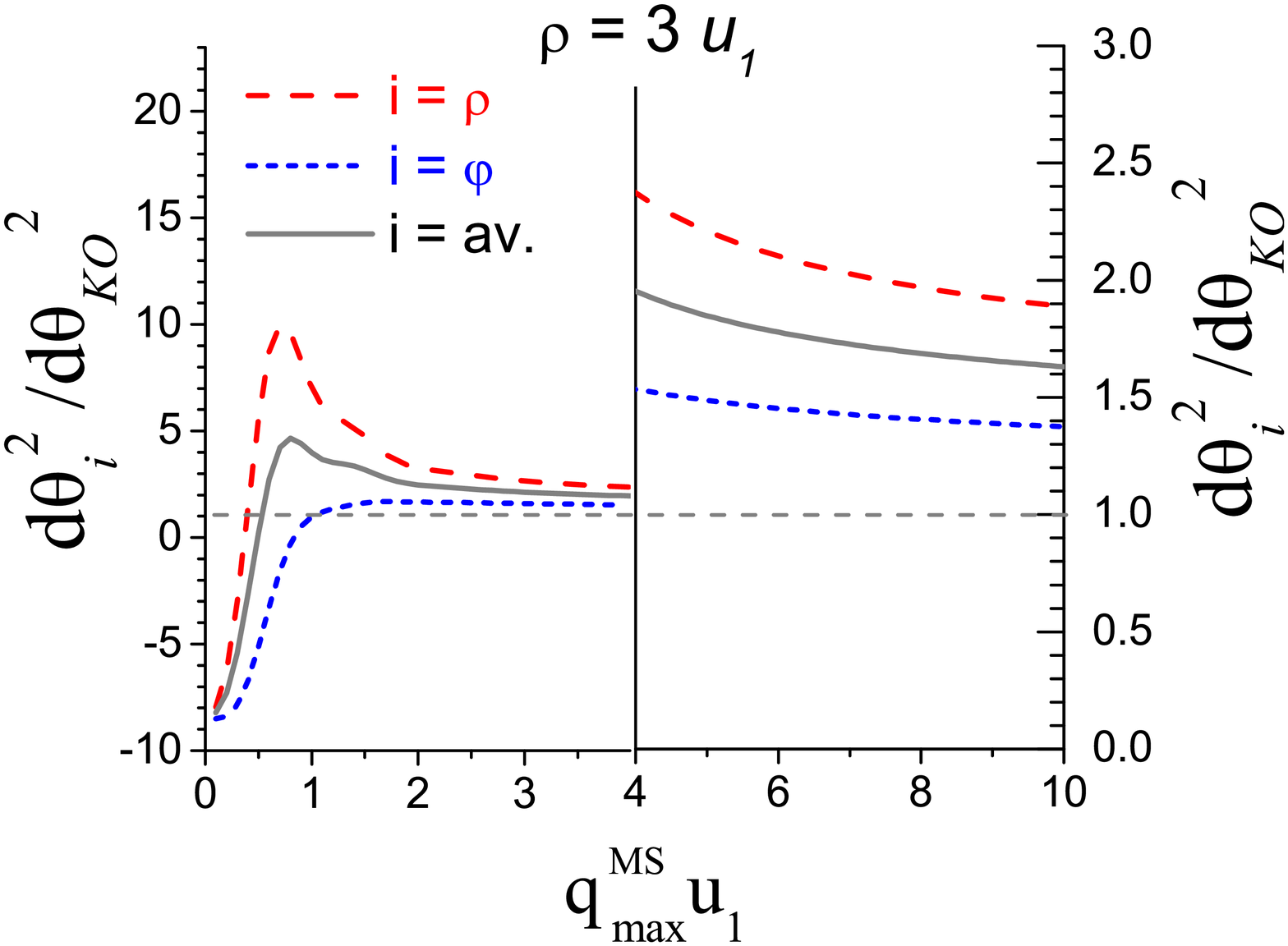}}\\
\caption{Mean square angles of the scattering in both radial ($i =
\rho$) and azimuthal ($i = \varphi$) directions along with their
average ($i = av.$) versus the Eq. (12) momentum limit
$q_{max}^{MS}$ at $\rho = 0$ (top left), $\rho = 2 u_1$ (top
right), $\rho = 2.5 u_1$ (bottom left) and $\rho = 3 u_1$ (bottom
right), plotted as a ratio to the analogous positive values,
evaluated using KO anzats (1). }
\end{center}
\end{figure}

Figs. 4 and 6 reveal considerably different transferred momentum
dependence at small $\rho < 2 u_1$ and large $\rho > 2 u_1$
distances. Indeed, the asymptote (10) demonstrates that Wigner
function contains both the nuclear density proportional, and the
cosine-dependent disproportionate parts, describing, respectively,
the local, high momentum and the distant, interference-sensitive,
low-momentum transfers. The interference effects, naturally,
induce Wigner function (9) oscillations at $\rho > a u_1$, where,
say, $a > 1.5$, demonstrating the distant action of the region
$\rho \sim a u_1 - b$ of the high nuclear density $n_n(a u_1 - b)
\gg n_n(a u_1)$, becoming pronounced at collision parameters $u_1
\leq b \leq R$. The same interference part of Eq. (9) also gives
rise to the strong scattering asymmetry in the region $2 \div 3
u_1$ of about 10 \% of the nuclei.

These two qualitatively different Eq. (10) parts correspond to
Kitagawa-Ohtsuki \cite{kit} and Lindhard \cite{lin} diffusion
coefficients, respectively. Despite the former neglects the
incoherent scattering reduction by correlations \cite{maz2,ter,
baz, lju, art, tik}, it describes the incoherent scattering in the
dense nuclear region with an accuracy of 5-10 \%. However the
strong nuclear density (2) decrease at $\rho > 2 u_1$ makes the
Kitagawa-Ohtsuki ansatz (1) inapplicable \cite{nit, nit2} for
treating the ultimate channeling stability problem for which both
Lindhard's transverse energy diffusion and electron scattering are
primarily important. The free from introduction of any approximate
limit local nuclear dechanneling treatment is provided by Eqs. (8)
and (9), taking into consideration all the peculiarities of both
large-angle single and small-angle multiple scattering addressed
bolow.

\subsection{The necessity of considerarion and relative role of single and multiple scattering in simulations}

Incoherent scattering treatment is crucial for both accuracy and
efficiency of simulations of particle propagation through
crystals. However, the really polar approaches of simulating all
the incoherent channeling effects by using solely the mean square
angle definition \cite{kit,tar,sca2} on the one hand, and of the
sampling successive single scatterings either classically
\cite{rob,sol} or quantumly \cite{tik3,art} on the other, coexist
in the literature for decades.

Our approach \cite{tik3}, verified in
\cite{ban,ban2,maz,syt,bar3,ban3,gui,tik5} and other
investigations, includes the features of both of them, combining
the sampling of both small angle multiple and large angle single
scattering, the latter of which is most consonant to \cite{art} in
the necessity of its quantum treatment. The introduced Eq. (9)
delivers both the firm grounds and calculation capabilities to the
method \cite{tik3}. However, before going to the latter, let us
remind the peculiarities of multiple Coulomb scattering theory
\cite{wil,mol,bet} application to the simulations of particle
propagation in crystals, which are still under discussion.

\begin{figure}
 \begin{center}
\resizebox{70mm}{!}{\includegraphics{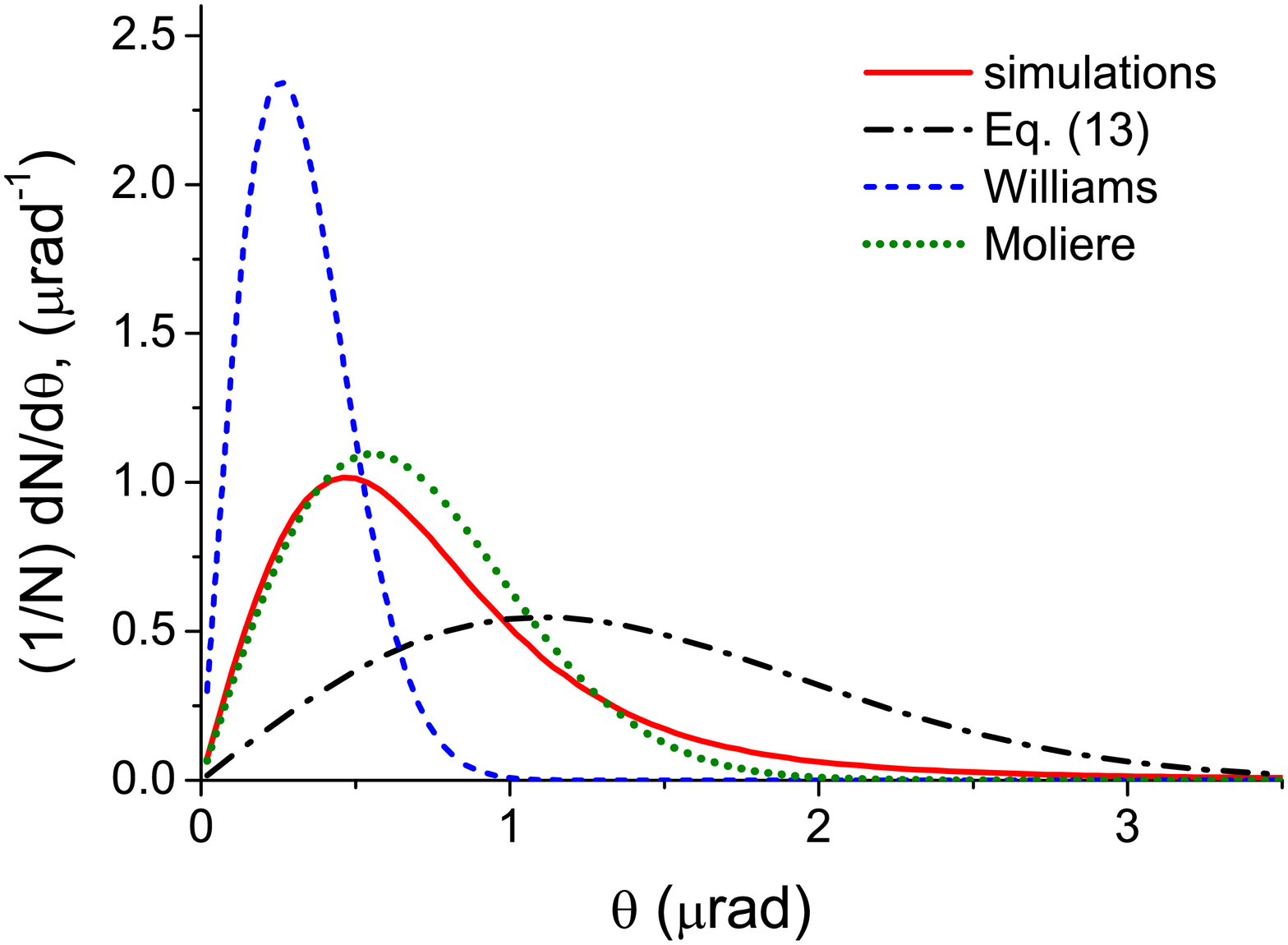}}
\hspace{1cm}\resizebox{70mm}{!}{\includegraphics{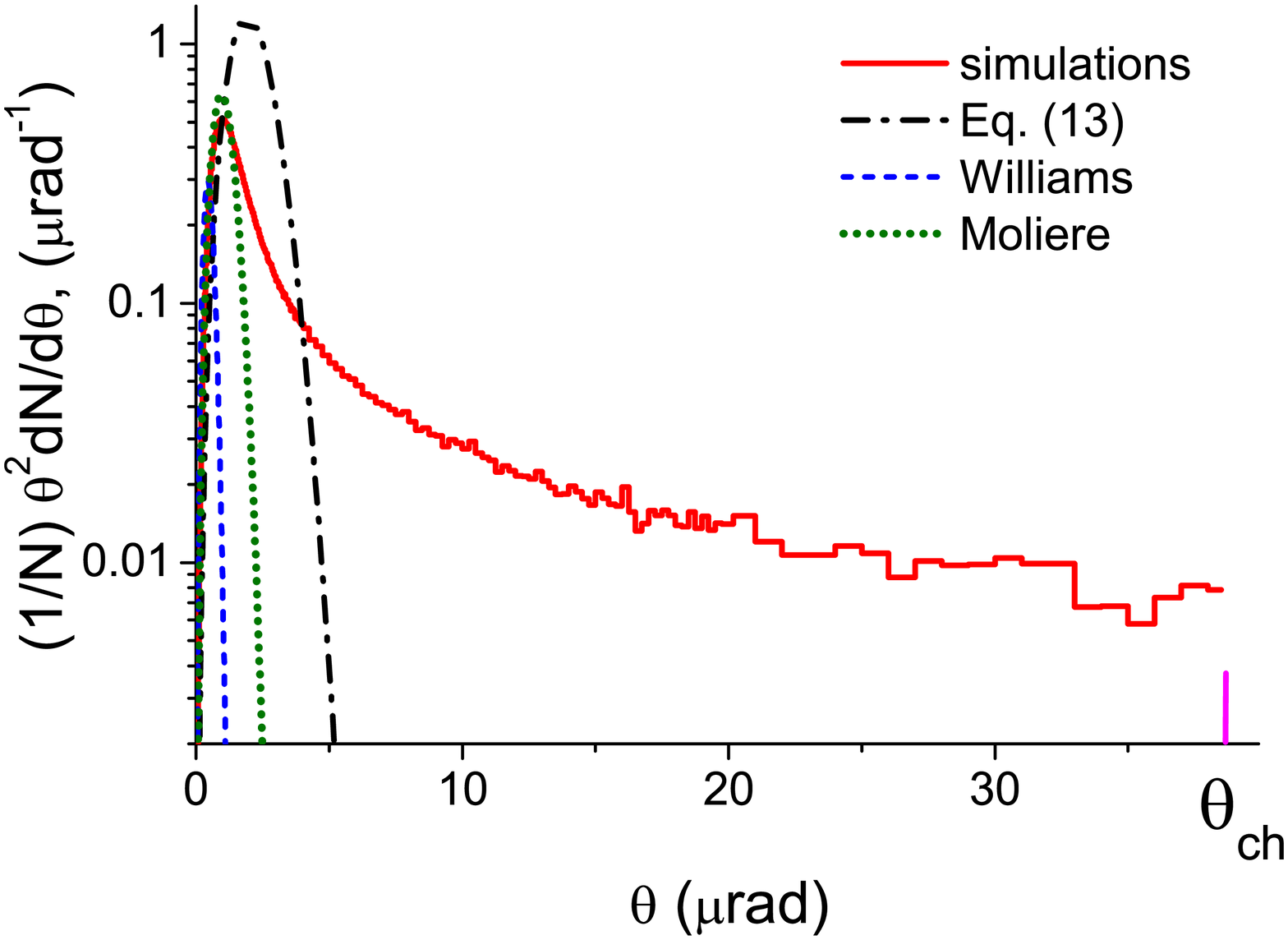}}
 \caption{Particle number (left) and deflection angle square (right) distributions
 of 150 GeV electrons, scattered within a typical trajectory
 simulation step (17): simulated directly (solid, red) and
 evaluated using Gaussian polar angle distribution with mean
 square angles determined by the usual (13) (dash-dot, black),
 Williams's (dashed, blue) and Moliere's (dotted, green lines) formulae.}
\end{center}
\end{figure}

Though multiple Coulomb scattering theory \cite{wil,mol,bet} (see
also \cite{jac,ter,sig}) is thoroughly developed and widely tested
experimentally, its formal application to the channeling
simulations is highly questionable \cite{tik3}. First of all,
remind that the former predicts a nearly Gaussian angular
distribution, characterized by the mean square angle
\begin {equation} % 13
\theta_s^2(l) = 8 \pi n \frac{Z^2 \alpha^2}{p^2 v^2}
\ln\left(\frac{\theta_{max}}{\theta_{min}}\right) l = 16 \pi n
\frac{Z^2 \alpha^2}{p^2 v^2} \ln\left(\frac{204}{Z^{1/3}}\right)l,
\end{equation}
where $n$ is a nuclear number density and $l$ a scattering length.
We consider here only the ultra-relativistic small angle limit and
adopt the slightly arbitrary numerical coefficients from
\cite{jac}.  Proceeding from the quantum principles, the theory
\cite{wil,mol,bet} introduces both the lower
\begin {equation} % 14
\theta_{min} = \frac{Z^{1/3}}{192} \frac{m}{p},
\end{equation}
where $m$ and $p$ are particle mass and momentum, and the upper
\begin {equation} % 15
\theta_{max} = \frac{274}{Z^{1/3}} \frac{m}{p}
\end{equation}
scattering angle limits for high energy particles. Essential point
of both \cite{baz,lju,tik,art} and the present paper is the widely
discussed above need of redefinition of the lower integration
limit (14) in the presence of coherent scattering in crystals.

However, the upper one (15), being often formally used with both
the mean square angle (13) and Gaussian distribution, is also
mostly inapplicable to the trajectory simulations in crystals
\cite{tik3}. Instead of the coherent scattering effect at the
small scattering angles, Eq. (13) inapplicability at the large
ones is related with the unaccustomed short length of trajectory
simulation steps, the case of which finds relatively specific
complex treatment in Moliere scattering theory
\cite{mol,bet,jac,sig}. The latter elucidates that single
scattering does not play considerable role if only $2.5~ \theta_s
> \theta_{max}$, as in a target with many radiation lengths
thickness. However, if $2.5~\theta_s < \theta_{max}$, the Gaussian
distribution is applicable solely at the angles $\theta < 2.5~
\theta_s$, while at $2.5~ \theta_s <\theta < \theta_{max}$ single
Coulomb scattering dominates.

The point is that the submicron length scale of simulated particle
trajectories in crystals results in the root mean square angles
$\theta_s$, being drastically smaller than the maximal angle
$\theta_{max}$ (15) and making both the single scattering
essential and the application of the angle $\theta_{max}$ as an
upper limit in Eq. (13) highly inconsistent. Indeed, Eq. (13)
involves a very wide single scattering angle interval $\theta_s
\leq \theta \leq \theta_{max}$ into the evaluation of the width
$\theta_s$ of the Gaussian distribution, covering, at the same
time, only a much smaller interval $0 <\theta < \theta_s$, which
does not include the angles $\theta_s \ll \theta \leq
\theta_{max}$, quite contradictory involved in $\theta_s$
evaluation through Eq. (15) at the same time. That is why, taken
with the upper limit (15),  Eq. (13) considerably overestimates
the mean square angle, making the Gaussian unphysically wide --
see Fig. 7. The book \cite{jac} elucidates accordingly that at
less than 200 collisions the true distribution, simulated here by
an elemental Monte Carlo for Fig. 7, "is more sharply peaked at
zero angle than a Gaussian" \cite{jac,sig}, as one can indeed see
in Fig. 7. That is why all the single scattering angles $\theta_s
\leq \theta \leq \theta_{max}$ should be disregarded in the mean
scattering angle evaluation by redefining the upper limit in (15)
according to the implicit condition $\theta_{max} \simeq
\theta_s(\theta_{max})$ \cite{mol,bet,tik3}.

The first realistic estimate \cite{wil}
\begin {equation} % 16
\theta_c = \frac{Z \alpha}{p v} \sqrt{4 \pi n l}
\end{equation}
of the actual width of the Gaussian distribution was introduced by
equating the scattering probability at the angles $\theta >
\theta_c$ to unity. However, the value (16), in fact,
underestimates the angular distribution width (see Fig. 7), since
a few scatterings by the angles $\theta < \theta_c$ often result
in a multiple scattering angle $\theta > \theta_c$. The correct
estimate $B\theta_c$, which contained a coefficient $B$,
determined from some implicit condition, was finally introduced in
\cite{mol,bet} along with the involved formula for the more exact
angular distribution.

To avoid the usage of the latter, we suggested \cite{tik3} to
sample both the small angle multiple and large angle single
scattering jointly within each trajectory step. At that, the
minimal angle of single scattering can be chosen to be either
equal or smaller than the root mean square multiple scattering
angle, evaluated using an estimate of the same as the upper
integration limit. This seemingly loose choice of the latter is,
in fact, validated by the adjustment of the single scattering
process, self-adapting to the choice of the minimal single
scattering angle. The multiple scattering simulation is, in fact,
optional for this method, being used to avoid the simulation of
some number of single scattering events. However, since the latter
is quite modest for the short trajectory steps, the multiple
scattering consideration can be reasonably abandoned in favor of
the single scattering simulations with the properly chosen lower
cutoff \cite{tik,tik3,art}.

To compare the different mean square angle definitions, we have
simulated the particle scattering in the screened Coulomb
potential by the angles $0 < \theta \leq \theta_{max} \simeq 300~
\mu rad$  for the 150 GeV energy used in the first experiment
\cite{sca3} on negatively charged particle multiple volume
reflection \cite{tik5}. Fig. 7, left, presents the simulated
angular distribution along with the three Gaussians, built for the
mean square angles determined, respectively, by Eqs. (13)-(15);
(16) and $\theta_s = B \theta_c$. All the four distributions were
evaluated for the typical trajectory step length
\begin {equation} % 17
\Delta l = 0.02~ \AA /\theta_{ch} \sim 0.05~\mu m.
\end{equation}
Fig. 7, left, demonstrates, that, being applied with the upper
limit (15), the definition (13) drastically overestimates the true
width of the angular distribution, correctly predicted by Moliere
theory and reproduced by simulations.

At the same time, starting from $5 \mu rad$, the Gaussian
distribution with any mean square angle definition essentially
underestimates the single scattering distribution tail, which both
gives considerable contribution to the mean square angle up to the
maximum single scattering angle $\theta_{max} \simeq 300~\mu rad$
and determines the actual, generally stochastic appearance of the
scattering particle trajectories. Also, since $\theta_{ch} \sim 40
~\mu rad \gg 5 \mu rad$, the single scattering processes is
completely responsible for the effects of instant dechanneling and
rechanneling. At the same time, owing both to the fast cross
section decrease with angle and small trajectory steps length,
single scattering simulations are not time consuming, free from
the problem with the logarithm in the mean square angle formula
(13) and readily reproduce both the not completely Gaussian small
angle, the single scattering large angle, and the most difficult
to handle intermediate angle regions of Moliere angular
distribution, highly unappropriate for efficient sampling. By this
reason we have mostly relied in our simulations
\cite{ban,ban2,maz,syt,bar3,ban3,tik3,gui,tik5} on the single
scattering sampling, while the multiple scattering one was applied
only optionally to accelerate the simulation process.

Instead of the previously known average estimates, the above
consideration treats the effect of incoherent scattering reduction
in crystals \cite{maz2,ter,baz,lju,tik,art} quantitatively at any
arbitrary point in the impact parameter plane. Both Eqs. (8)-(12)
and Figs. 4 and 6 reveal the indispensable role of the multiple
scattering consideration describing its nontrivial characteristics
(12) in the presence of collision correlations. Indeed, Fig. 4
highlights the impossibility to introduce the positive small angle
scattering probability or cross section. However, according to
Fig. 6, the multiple scattering angles (12) can be made positive
by the high enough integration limit $q_{max}^{MS}$ choice, in
order to be appropriate for scattering angle sampling. Figs. 4, 6
demonstrate that both scattering probability (9) and mean square
angles (12) becomes positive starting from the rather small
momentum transfer $q_{max}^{MS} \simeq \hbar /u_1$.

The most direct way to consistently combine single scattering with
multiple one within an arbitrary trajectory step is to simulate
the "small" angle scattering as a cumulative multiple process of
all the momentum transfers $q < q_{max}^{MS}$ using the mean angle
squares (12) and to perform further the single scattering sampling
using the local probability (9), being positive at $q_{max}^{MS} <
q < \theta_{max} p$.

However, being the most straightforward, this approach includes
two relatively laborious step of both (9) and (12) evaluation, the
former of which can be avoided, including the difference of Eq.
(9) predictions from the ones of Kitagawa-Ohtsuki ansatz into the
integrals of Eq. (12) type.

According to the central limit theorem a loose choice of the
momentum transfer limit $q_{max}^{MS}$ will have negligible
influence on the simulated angular distribution, provided both the
small $q_{max}^{MS}$ value and thoroughly evaluated mean squared
angles (12) use.

The above consideration, undertaken here for the axial case, can
be extend to the planar one by either a direct derivation,
analogous to Eqs. (3)-(8), or through the averaging of Eqs.
(9)-(12) along straight trajectories traversing atomic strings. If
to picture crystal planes as the ones being assembled from atomic
strings, it becomes clear, that the scattering intensity increase
in the radial direction, illustrated by both Figs. 4 and 6, has to
result in the same of the nuclear dechanneling rate in the region
of $2.5~u_1 \sim 0.2 $\AA, critical for positively charged
particle dechanneling.

Besides the numerous particle beam manipulation problems, the
developed approach can be applied to refine both the radiation and
pair production simulation methods of Refs. \cite{gui, bar3, ban,
ban3, ban2}.

\section{Conclusions}

A theory of incoherent particle scattering in oriented crystals in
the high energy limit, in which relativistic particle motion in
the averaged field of atomic strings and planes is mostly
classical, is developed. Quantum interference effects of
scattering by the atomic cores along a classical trajectory are
treated through applying the Wigner function, determined in the
phase space of transverse particle motion. The recently observed
effect of incoherent scattering reduction in the presence of atom
distribution inhomogeneity is, for the first time, described
precisely on a classical particle trajectory. An impossibility to
introduce the local small-angle scattering probability is revealed
and a novel definition of mean square scattering angle is
introduced to incorporate the quantum scattering features into the
classical trajectory simulations. A considerable excess over the
Kitagawa-Ohtsuki diffusion coefficient is numerically demonstrated
in the region of negligible nuclear density. In general, present
theory makes it possible to refine a numerical treatment of any
experiment on high energy particle scattering and radiation in
crystals ever conducted or planned.

\subsection{Asknowledgments}

Financial support by the European Commission through the PEARL
Project, GA 690991, is gratefully acknowledged.

The author is obliged to both the anonymous referee and X. Artru
for sharing the understanding of importance of the treated
problem.

\end{document}